\pdfoutput=1

\RequirePackage{fix-cm}

\documentclass[twocolumn]{svjour3}

\smartqed

\usepackage{graphicx}
\usepackage{amssymb,amsmath}
\usepackage{txfonts}
\usepackage{float}
\usepackage[numbers,sort&compress]{natbib}
\usepackage{hyperref}

\hypersetup{pdftitle = Escape dynamics in the H\'{e}non-Heiles Hamiltonian system, pdfauthor = Euaggelos E. Zotos, pdfsubject= Escape dynamics, pdfkeywords = Hamiltonians Escapes Fractals}
\hypersetup{colorlinks = true, linkcolor = blue, anchorcolor = red, citecolor = blue, filecolor = red, pagecolor = red, urlcolor = blue}

\journalname{Meccanica}

\begin{document}

\title{An overview of the escape dynamics in the H\'{e}non-Heiles Hamiltonian system}

\author{Euaggelos E. Zotos}

\institute{Department of Physics, School of Science, \\
Aristotle University of Thessaloniki, \\
GR-541 24, Thessaloniki, Greece \\
Corresponding author's email: {evzotos@physics.auth.gr}}

\date{Received: 14 November 2016 / Accepted: 24 February 2017 / Published online: 3 March 2017}

\titlerunning{An overview of the escape dynamics in the H\'{e}non-Heiles Hamiltonian system}

\authorrunning{Euaggelos E. Zotos}

\maketitle

\begin{abstract}

The aim of this work is to revise but also explore even further the escape dynamics in the H\'{e}non-Heiles system. In particular, we conduct a thorough and systematic numerical investigation distinguishing between trapped (ordered and chaotic) and escaping orbits, considering only unbounded motion for several energy levels. It is of particular interest, to locate the basins of escape towards the different escape channels and relate them with the corresponding escape periods of the orbits. In order to elucidate the escape process we conduct a thorough investigation in several types of two-dimensional planes and also in a three-dimensional subspace of the entire four-dimensional phase space. We classify extensive samples of orbits by integrating numerically the equations of motion as well as the variational equations. In an attempt to determine the regular or chaotic nature of trapped motion, we apply the SALI method, as an accurate chaos detector. It was found, that in all studied cases regions of non-escaping orbits coexist with several basins of escape. Most of the current outcomes have been compared with previous related work.


\end{abstract}

\section{Introduction}
\label{intro}

Escaping particles from dynamical systems is a subject to which has been devoted many studies over the years. Especially the issue of escapes in Hamiltonian systems is directly related to the problem of chaotic scattering which has been an active field of research over the last decades and it still remains open (e.g., \cite{BTS96,BST98,BGOB88,BOG89,BGO90,CPR75,C90,CK92,E88,JS88,JT91,JLS99,ML02,NH01,OT93,PH86,SASL06,SSL07,SS08,SHSL09,SS10}). It is well known, that particular types of Hamiltonian systems have a finite energy of escape and for lower values of the energy the equipotential surfaces of the systems are closed and therefore escape is impossible. For energy levels beyond the escape energy however, these surfaces open creating exit channels through which the particles can escape to infinity. The literature is replete with studies of such ``open" Hamiltonian systems (e.g., \cite{BBS09,CKK93,KSCD99,NH01,SHA03,STN02,SCK95,SKCD95,SKCD96,Z14a,Z14b,Z15a,Z15b}).

Usually, the infinity acts as an attractor for an escape particle, which may escape through different channels (exits) on the equipotential curve or on the equipotential surface depending whether the dynamical system has two or three degrees of freedom, respectively. Therefore, it is quite possible to obtain basins of escape, similar to basins of attraction in dissipative systems or even the Newton-Raphson fractal structures. Basins of escape have been studied in several papers (e.g., \cite{AVS01,AVS09,BBS08,BBS09,Z14a,Z14b,Z15a,Z15b}). The reader can find more details regarding basins of escape in \cite{C02}.

The H\'{e}non-Heiles potential is undoubtedly one of the most simple, classical and characteristic example of open Hamiltonian systems with two degrees of freedom. A huge load of research has been devoted on this dynamical system (e.g., \cite{W84,CTW82,F91,RGC93,dML99,AVS01,AS03,AVS03,B05,CMV05,SASL06,SSL07,BBS08,SS08,AVS09,BBS09,SHSL09,SS10,BSBS12,BSBS14,CSS13}). At this point we should emphasize, that all the above-mentioned references regarding previous studies in the H\'{e}non-Heiles system are exemplary rather than exhaustive, taking into account that a large quantity of related literature exists. The vast majority of these papers deals mostly either with the discrimination between regular and chaotic motion or with the escape properties of orbits. In the present paper, which can be considered as a mini-review paper, we present a recollection of interesting aspects of the H\'{e}non-Heiles system that have been discussed in previously published papers. However, in this paper we shall proceed one step further by classifying, for the first time, initial conditions of orbits in the 3D subspace of the entire 4D phase space.

The structure of the article is as follows: In Section \ref{mod} we present a detailed description of the properties of the H\'{e}non-Heiles system. All the different computational methods used in order to determine the character (ordered vs. chaotic and trapped vs. escaping) of the orbits are described in Section \ref{cometh}. In the following Section, we conduct a thorough numerical analysis of several sets of initial conditions of orbits presenting in detail all the results of our computations. Our article ends with Section \ref{disc}, where the discussion is given.

\section{Properties of the H\'{e}non-Heiles system}
\label{mod}

The corresponding potential of the H\'{e}non-Heiles system \cite{HH64} is given by
\begin{equation}
V(x,y) = \frac{1}{2} \left(x^2 + y^2 \right) + x^2 y - \frac{1}{3}y^3.
\label{HHpot}
\end{equation}
It can be seen in Eq. (\ref{HHpot}) that the potential is in fact composed of two harmonic oscillators that have been coupled by the perturbation terms $x^2 y - 1/3 y^3$.

The H\'{e}non-Heiles potential along with the dihedral $D4$ potential \cite{AGK89} and the Toda potential \cite{T67} belong to a specialized category of potentials. In the pioneering spirit of H\'{e}non and Heiles, the aim was to select potentials which are generic in their basic properties, but convenient computationally, so that large numbers of computations could be performed. Furthermore, the H\'{e}non-Heiles potential admits a discrete triangular rotation $D3$ $(2\pi/3)$ symmetry.

The basic equations of motion for a test particle with a unit mass $(m = 1)$ are
\begin{align}
\ddot{x} &= - \frac{\partial V}{\partial x} = - x - 2xy, \nonumber\\
\ddot{y} &= - \frac{\partial V}{\partial y} = - x^2 - y + y^2,
\label{eqmot}
\end{align}
where, as usual, the dot indicates derivative with respect to the time.

\begin{figure*}[!t]
\centering
\resizebox{\hsize}{!}{\includegraphics{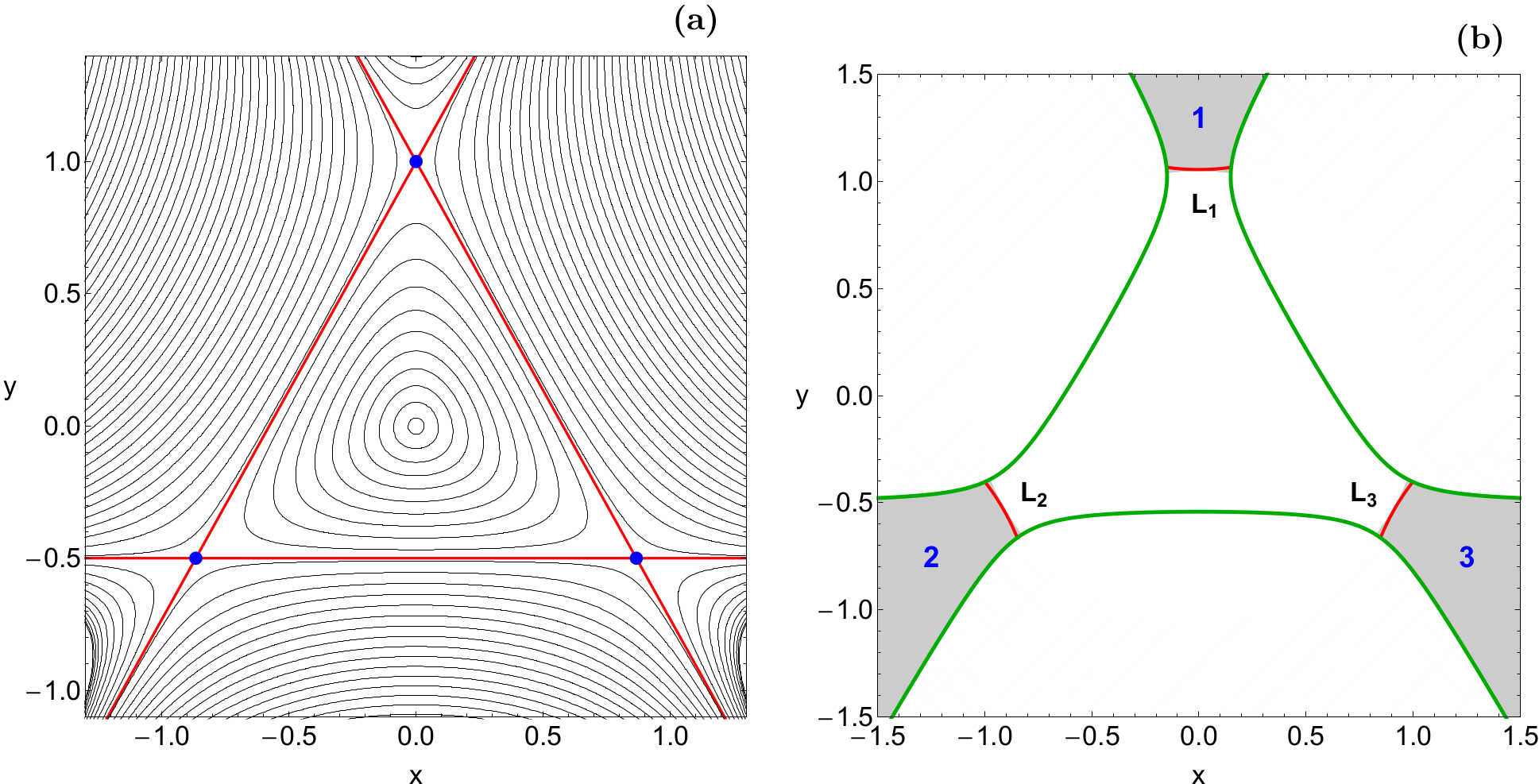}}
\caption{(a-left): Equipotential curves of the H\'{e}non-Heiles potential for various values of the energy $h$. The equipotential curve corresponding to the energy of escape $h_{esc}$ is shown in red, while the positions of the three saddle points are marked by blue dots; (b-right): The open ZVC on the configuration $(x,y)$ plane when $h = 0.20$. $L_1$, $L_2$ and $L_3$ indicate the three unstable Lyapunov orbits plotted in red.}
\label{pot}
\end{figure*}

Consequently, the Hamiltonian to potential (\ref{HHpot}) reads
\begin{equation}
H = \frac{1}{2}\left(\dot{x}^2 + \dot{y}^2 + x^2 + y^2\right) + x^2 y - \frac{1}{3}y^3 = h,
\label{ham}
\end{equation}
where $\dot{x}$ and $\dot{y}$ are the momenta per unit mass conjugate to $x$ and $y$, respectively, while $h > 0$ is the numerical value of the Hamiltonian, which is conserved. It is seen that the Hamiltonian is symmetric with respect to $x \rightarrow - x$, while $H$ also manifests a $2\pi/3$ rotation symmetry.

The H\'{e}non-Heiles potential has two degrees of freedom (2-dof) and a finite energy of escape $(h_{esc})$ which is equal to 1/6. For values of energy $h < h_{esc}$, the equipotential curves of the system are closed thus making motion inescapable. However, for larger energy levels $(h > h_{esc})$, the equipotential curves open and three exit channels appear through which the test particles may escape to infinity. The equipotential curves of the H\'{e}non-Heiles potential for various values of the energy $h$ are shown in Fig. \ref{pot}a. In the same plot, the equipotential corresponding to the energy of escape $h_{esc}$ is plotted with red color. Furthermore, the potential has a stable equilibrium point at $(x,y) = (0,0)$ and three saddle points: $(x,y)$ = $\{(0,1), (-\sqrt{3}/2 ,−1/2), (\sqrt{3}/2 ,−1/2)\}$. The saddle points constitute the three corners of the equipotential curve $V(x,y) = 1/6$, that can be seen with blue dots in Fig. \ref{pot}a. This triangular area that is bounded by the equipotential curve with energy $h = 1/6$ is called the ``interior region", while all the outside available area of motion is known as the ``exterior region". The open Zero Velocity Curve (ZVC) at the configuration $(x,y)$ plane when $h = 0.20 > h_{esc}$ is presented with green color in Fig. \ref{pot}b and the three channels of escape are shown.

An issue of paramount importance is the determination of the position as well as the time at which an orbit escapes. An open ZVC consists of several branches forming channels through which an orbit can escape to infinity. At every opening there is a highly unstable periodic orbit close to the line of maximum potential \cite{C79} which is called a Lyapunov orbit. Such an orbit reaches the ZVC, on both sides of the opening and returns along the same path, thus connecting two opposite branches of the ZVC. Lyapunov orbits are very important for the escapes from the system, since if an orbit crosses any one of these orbits with velocity pointing outwards moves always outwards and eventually escapes from the system without any further crosses with the surface of section (e.g., \cite{C90}). The passage of orbits through the Lyapunov orbits and their subsequent escape to infinity is the most conspicuous aspect of the transport, but crucial features of the bulk flow, especially at late times, appear to be controlled by diffusion through cantori which can trap orbits far vary long time periods. In Fig. \ref{pot}b, we denote the three unstable Lyapunov orbits by $L_i$, $i = 1, 2, 3$ using red color.

\section{Description of the computational methods}
\label{cometh}

In order to explore the orbital structure of the H\'{e}non-Heiles system, we need to define samples of orbits whose properties (regularity/chaos and trapped/escape) will be identified. For the 2D system, we define for each energy level (all tested energy levels are above the escape energy), dense uniform grids of $1024 \times 1024$ initial conditions. Our investigation takes place both in the configuration $(x,y)$ and also in the phase $(y,\dot{y})$ space for a better understanding of the escape mechanism. With polar coordinates $(r,\phi)$ in the origin of the coordinates $(0,0)$, the condition $\dot{r} = 0$ defines a two-dimensional surface of section, with two disjoint parts $\dot{\phi} < 0$ and $\dot{\phi} > 0$. Each of these two parts has a unique projection onto the configuration configuration $(x,y)$ space; we chose the $\dot{\phi} > 0$ part. For the phase $(y,\dot{y})$ space we consider orbits with initial conditions $(y_0, \dot{y_0})$ with $x_0 = 0$, while the initial value of $\dot{x_0}$ is obtained from the energy integral (\ref{ham}). For the 3D phase space a grid of $300 \times 300 \times 300$ initial conditions is defined inside the three-dimensional $(x,y,\dot{y})$ phase space.

For each initial condition, we integrated the equations of motion (\ref{eqmot}) as well as the variational equations using a double precision Bulirsch-Stoer \verb!FORTRAN 77! algorithm (e.g., \cite{PTVF92}) with a small time step of order of $10^{-2}$, which is sufficient enough for the desired accuracy of our computations (i.e., our results practically do not change by halving the time step). Our previous experience suggests that the Bulirsch-Stoer integrator is both faster and more accurate than a double precision Runge-Kutta-Fehlberg algorithm of order 7 with Cash-Karp coefficients. In all cases, the energy integral (Eq. (\ref{ham})) was conserved better than one part in $10^{-12}$, although for most orbits it was better than one part in $10^{-13}$.

The configuration and the phase space are divided into the escaping and non-escaping (trapped) space. Usually, the vast majority of the trapped space is occupied by initial conditions of regular orbits forming stability islands where a third integral is present. In many systems however, trapped chaotic orbits have also been observed (e.g., \cite{Z15b}). Therefore, we decided to distinguish between regular and chaotic trapped orbits. Over the years, several chaos indicators have been developed in order to determine the character of orbits. In our case, we chose to use the Smaller ALingment Index (SALI) method \cite{S01}, which has been proved a very fast, reliable and effective tool.

In our computations, we set $10^5$ time units as a maximum time of numerical integration. Our previous experience in this subject indicates, that usually orbits need considerable less time to find one of the exits in the limiting curve and eventually escape from the system (obviously, the numerical integration is effectively ended when an orbit passes through one of the escape channels and intersects one of the unstable Lyapunov orbits). Nevertheless, we decided to use such a vast integration time just to be sure that all orbits have enough time in order to escape. Remember, that there are the so called ``sticky orbits" which behave as regular ones and their true chaotic character is revealed only after long time intervals of numerical integration. A sticky orbit could be easily misclassified as regular by any chaos indicator if the total integration interval is too small, so that the orbit do not have enough time in order to reveal its true chaotic character. Here we should clarify, that orbits which do not escape after a numerical integration of $10^5$ time units are considered as non-escaping or trapped.

\section{Escape dynamics}
\label{numres}

Our main objective is to distinguish between trapped and escaping orbits for values of energy larger than the escape energy where the ZVCs are open and three channels of escape are present. Moreover, two additional properties of the orbits will be examined: (i) the directions or channels through which the particles escape and (ii) the time-scale of the escapes (we shall also use the term escape period). We decided to classify the initial conditions of orbits in both the configuration and the phase space into three main categories: (i) orbits that escape through one of the escape channels, (ii) non-escaping regular orbits and (iii) trapped chaotic orbits.

\subsection{Results for the 2D system}

Our exploration begins in the 2D phase space and particularly in the configuration $(x,y)$ space. We shall deal only with unbounded motion of test particles for values of energy in the set $h = \{0.17, 0.18, 0.19, 0.20, 0.22, 0.24, 0.26, 0.28, 0.30\}$.

\begin{figure*}[!t]
\centering
\resizebox{\hsize}{!}{\includegraphics{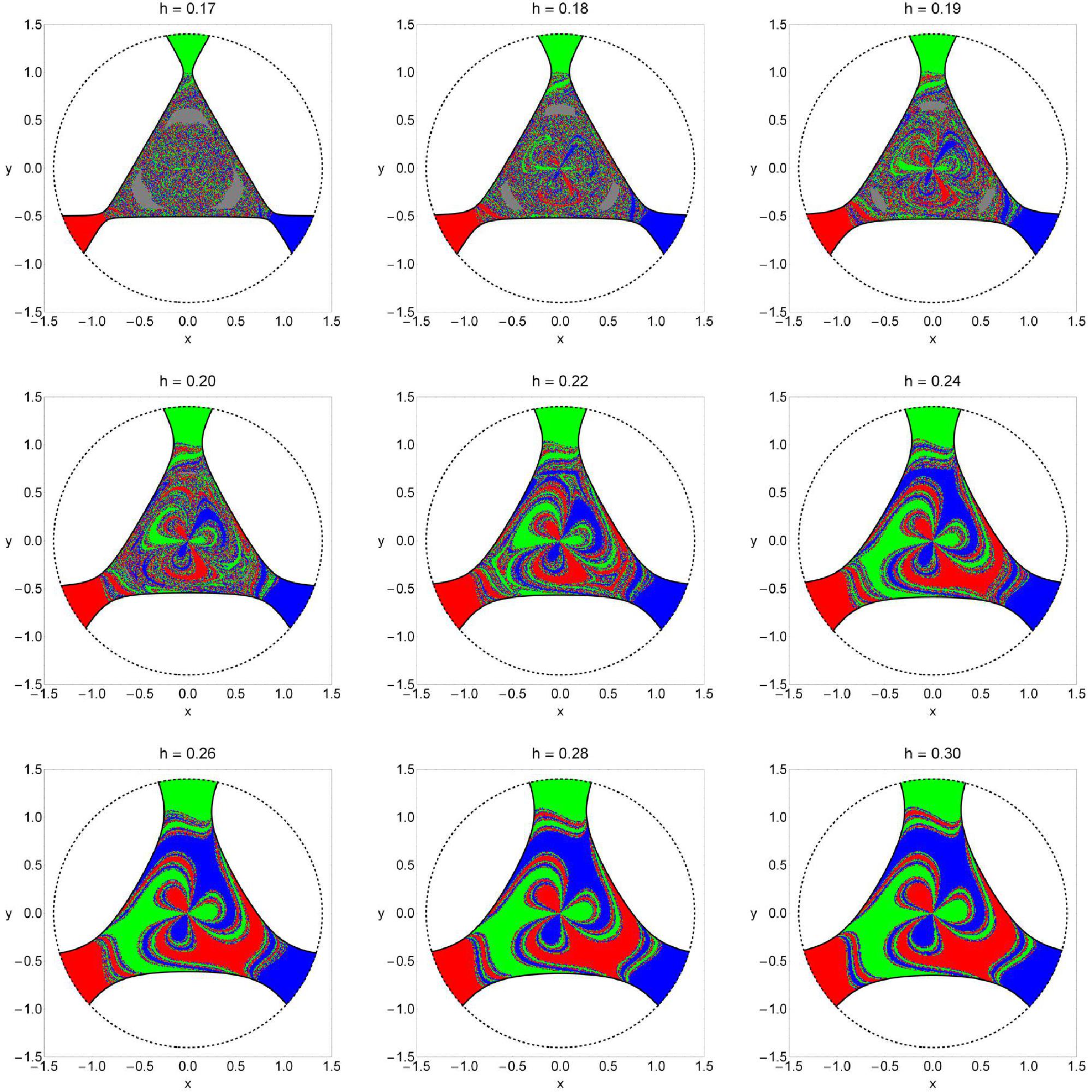}}
\caption{The structure of the $\dot{\phi} > 0$ part of the configuration $(x,y)$ space for several values of the energy $h$, distinguishing between different escape channels. The color code is as follows: Non-escaping regular (gray); trapped chaotic (black); escape through channel 1 (green); escape through channel 2 (red); escape through channel 3 (blue). The black, dashed circle denotes the scattering region.}
\label{xy}
\end{figure*}

\begin{figure*}[!t]
\centering
\resizebox{\hsize}{!}{\includegraphics{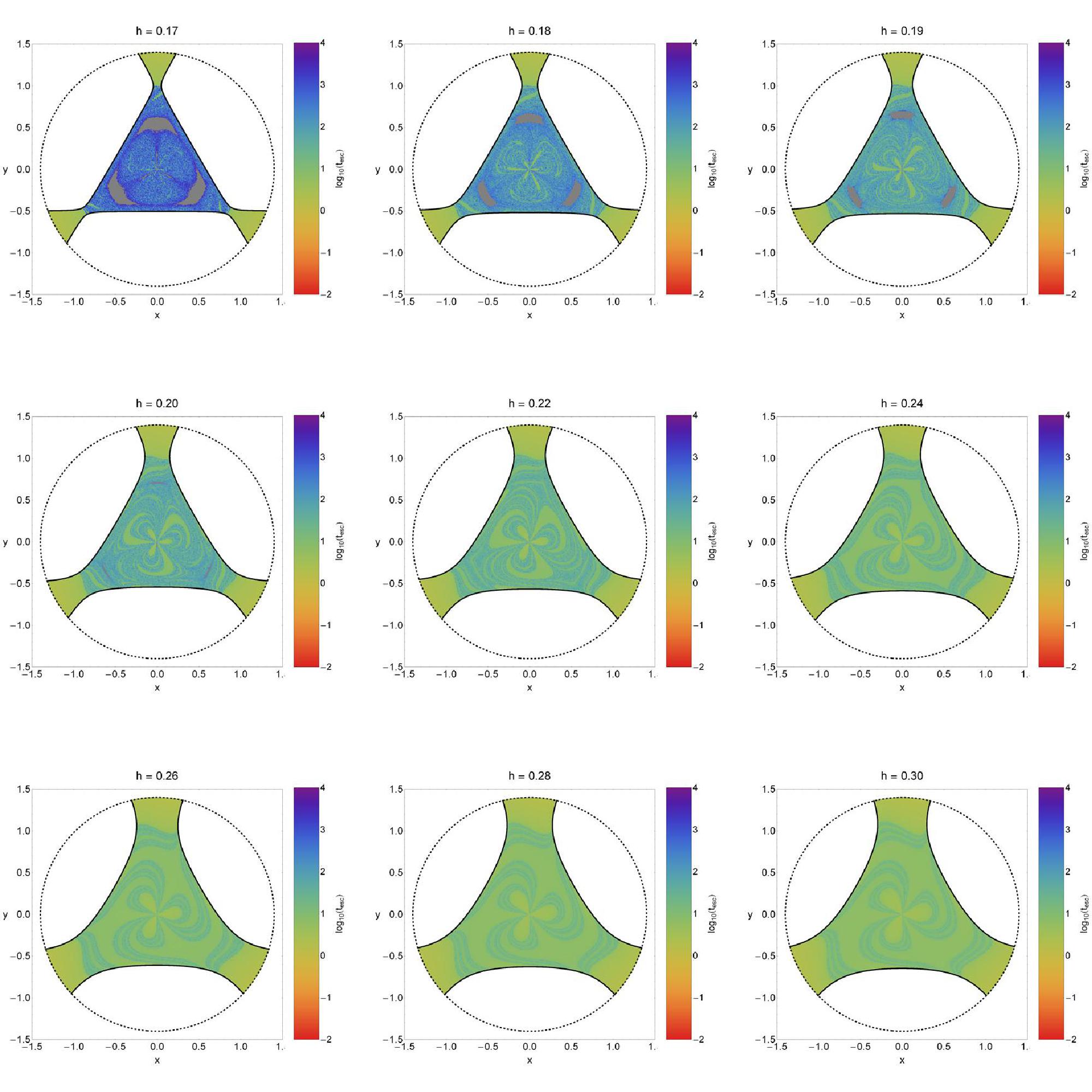}}
\caption{Distribution of the escape times $t_{\rm esc}$ of the orbits on the configuration $(x,y)$ plane. The darker the color, the larger the escape time. Trapped and non-escaping orbits are indicated by gray color.}
\label{xyt}
\end{figure*}

\subsubsection{The configuration $(x,y)$ space}

First of all, we will explore the escape process in the $\dot{\phi} > 0$ part of the configuration $(x,y)$ plane (the $\dot{\phi} < 0$ part gives similar results). Fig. \ref{xy} shows the structure of the $(x,y)$ plane for different values of the energy. Each initial condition is colored according to the escape channel through which the particular orbit escapes. The gray regions on the other hand, denote initial conditions of non-escaping regular orbits, while trapped chaotic orbits are indicated with black dots. The outermost black solid line is the Zero Velocity Curve (limiting curve) which is defined as $V(x,y) = h$. It is seen, that for values of energy larger but yet very close to the escape energy a substantial portion of the $(x,y)$ plane is covered by stability islands which correspond to initial conditions of non-escaping orbits surrounded by a very rich fractal structure. Looking carefully the grids we also observe that there is a highly sensitive dependence of the escape process on the initial conditions, that is, a slight change in the initial conditions makes the test particle escape through another channel, which is a classical indication of chaos. As the value of the energy increases the stability islands and the amount of trapped orbits is reduced and basins of escape emerge. These basins of escape are either broad regions or thin elongated spiral bands. Indeed, when $h = 0.30$ all the computed orbits of the grid escape and there is no indication of bounded motion or whatsoever. By the term basin of escape, we refer to a set of initial conditions that corresponds to a certain escape channel. The escape basins become smoother and more well-defined as the energy increases and the degree of fractality decreases\footnote{The fat-fractal exponent increases, approaching the value 1 which means no fractal geometry, when the energy of the system is high enough (e.g., \cite{BBS08}).}. The fractality is strongly related with the unpredictability in the evolution of a dynamical system. In our case, it can be interpreted that for high enough energy levels, the test particles escape very fast from the scattering region and therefore, the system's predictability increases.

The following Fig. \ref{xyt} shows how the escape times $t_{\rm esc}$ of orbits are distributed on the $(x,y)$ plane. Light reddish colors correspond to fast escaping orbits, dark blue/purple colors indicate large escape periods, while gray color denote both trapped and non-escaping orbits. We observe, that when $h = 0.17$, that is, a value of energy very close to the escape energy, the escape periods of the majority of orbits are huge corresponding to tens of thousands of time units. This however, is anticipated because in this case the width of the escape channels is very small and therefore, the orbits should spend much time inside the equipotential curve until they find one of the openings and eventually escape to infinity. As the value of the energy increases however, the escape channels become more and more wide leading to faster escaping orbits, which means that the escape period decreases rapidly. We found, that the longest escape rates correspond to initial conditions near the boundaries between the escape basins and near the vicinity of stability islands. On the other hand, the shortest escape periods have been measured for the regions without sensitive dependence on the initial conditions (basins of escape), that is, those far away from the fractal basin boundaries. This grid representation of the configuration plane gives us a much more clearer view of the orbital structure and especially about the non-escaping ordered orbits. In particular, we see that for $h = 0.2$ we have the last indication of stability islands, as for all higher energy levels studied all orbits escape, thus defying basins of escape. It should be pointed out that similar plots regarding the distribution of the escape times of orbits appear also in previous works (i.e., Figs. 3b and 4 in \cite{CSZS13}).

\subsubsection{The phase $(y,\dot{y})$ space}

\begin{figure*}[!t]
\centering
\resizebox{\hsize}{!}{\includegraphics{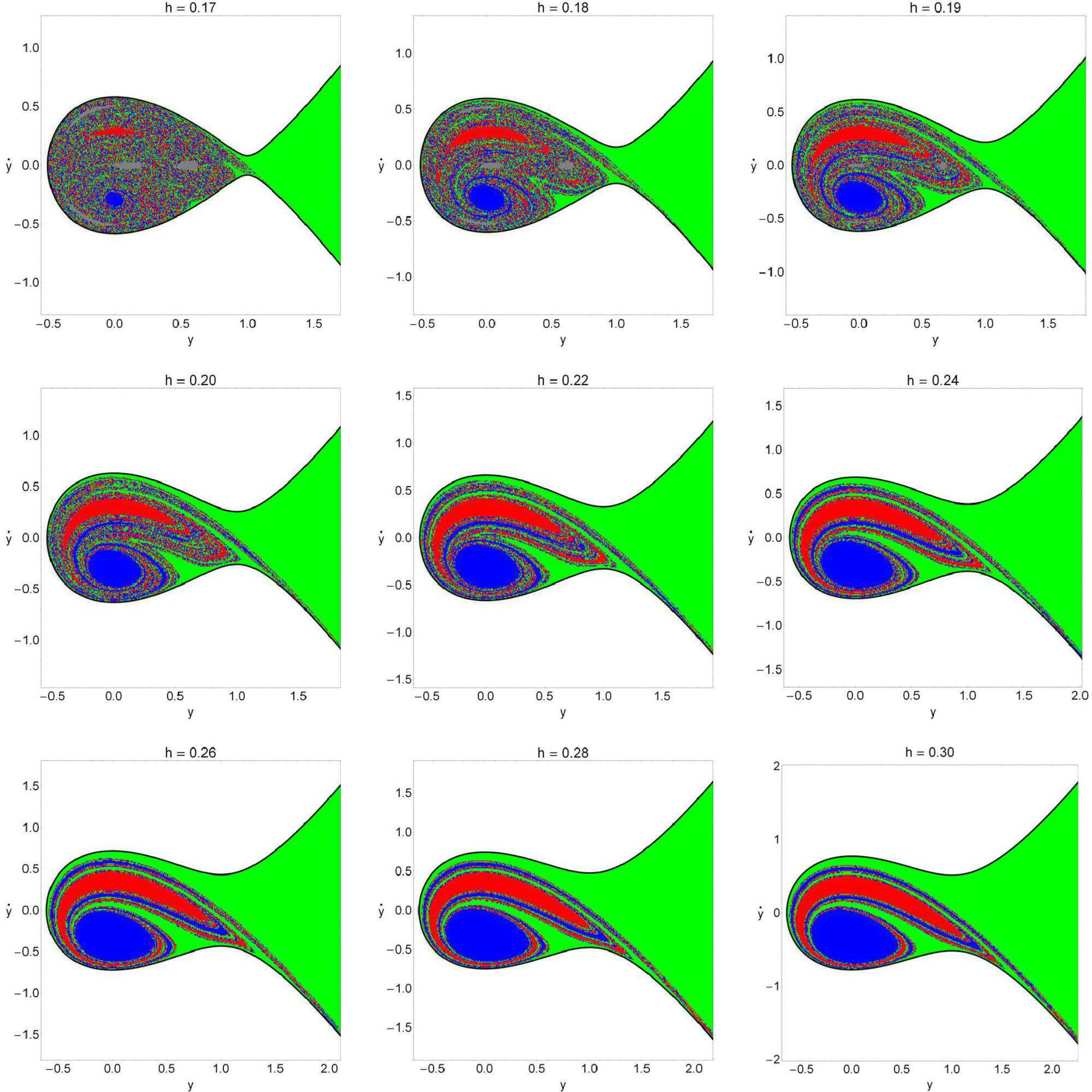}}
\caption{The structure of the phase $(y,\dot{y})$ space for several values of the energy $h$, distinguishing between different escape channels. The color code is as follows: Non-escaping regular (gray); trapped chaotic (black); escape through channel 1 (green); escape through channel 2 (red); escape through channel 3 (blue).}
\label{ypy}
\end{figure*}

\begin{figure*}[!t]
\centering
\resizebox{\hsize}{!}{\includegraphics{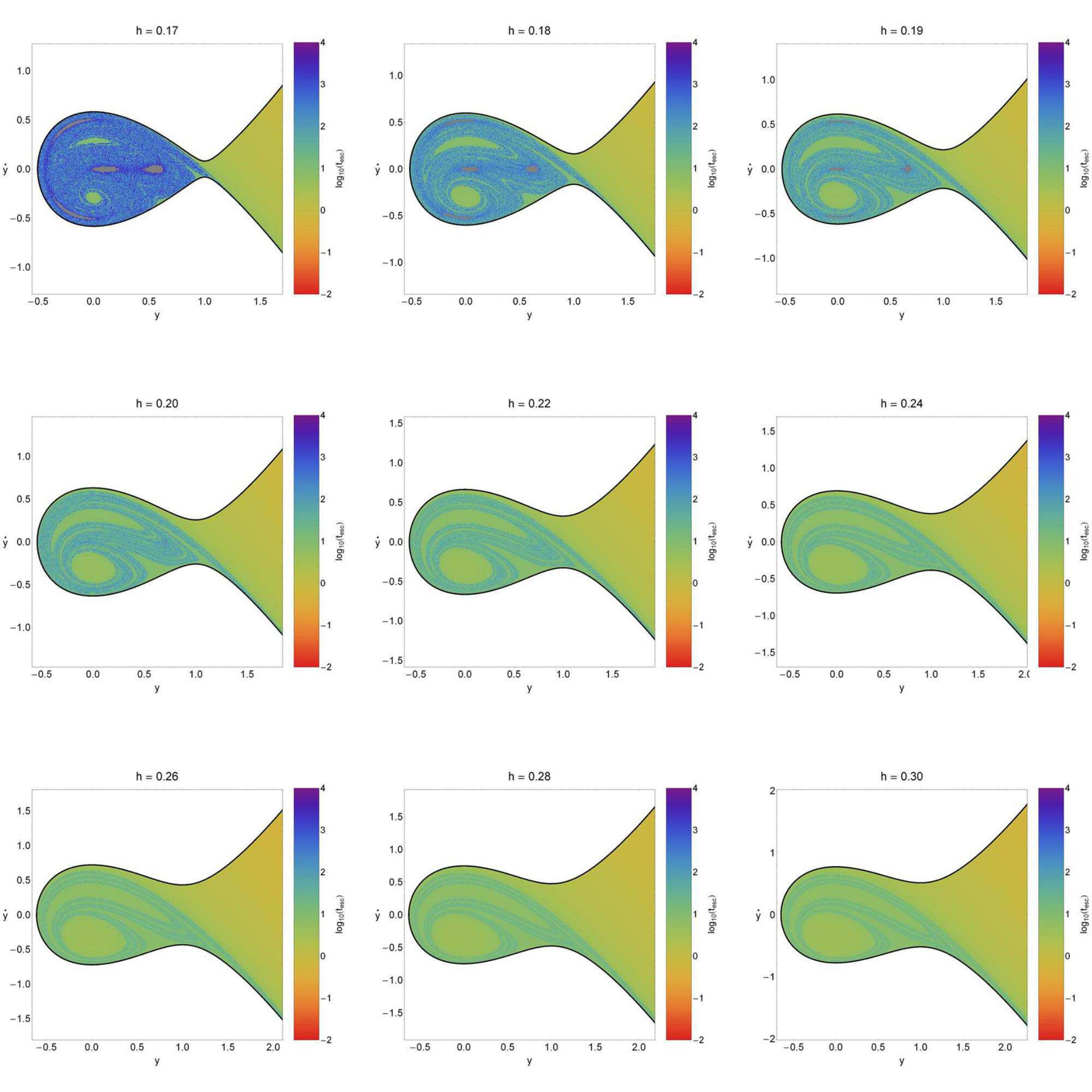}}
\caption{Distribution of the escape times $t_{\rm esc}$ of the orbits on the $(y,\dot{y})$ plane. The darker the color, the larger the escape time. Trapped and non-escaping orbits are indicated by gray color.}
\label{ypyt}
\end{figure*}

\begin{figure}[!t]
\begin{center}
\includegraphics[width=\hsize]{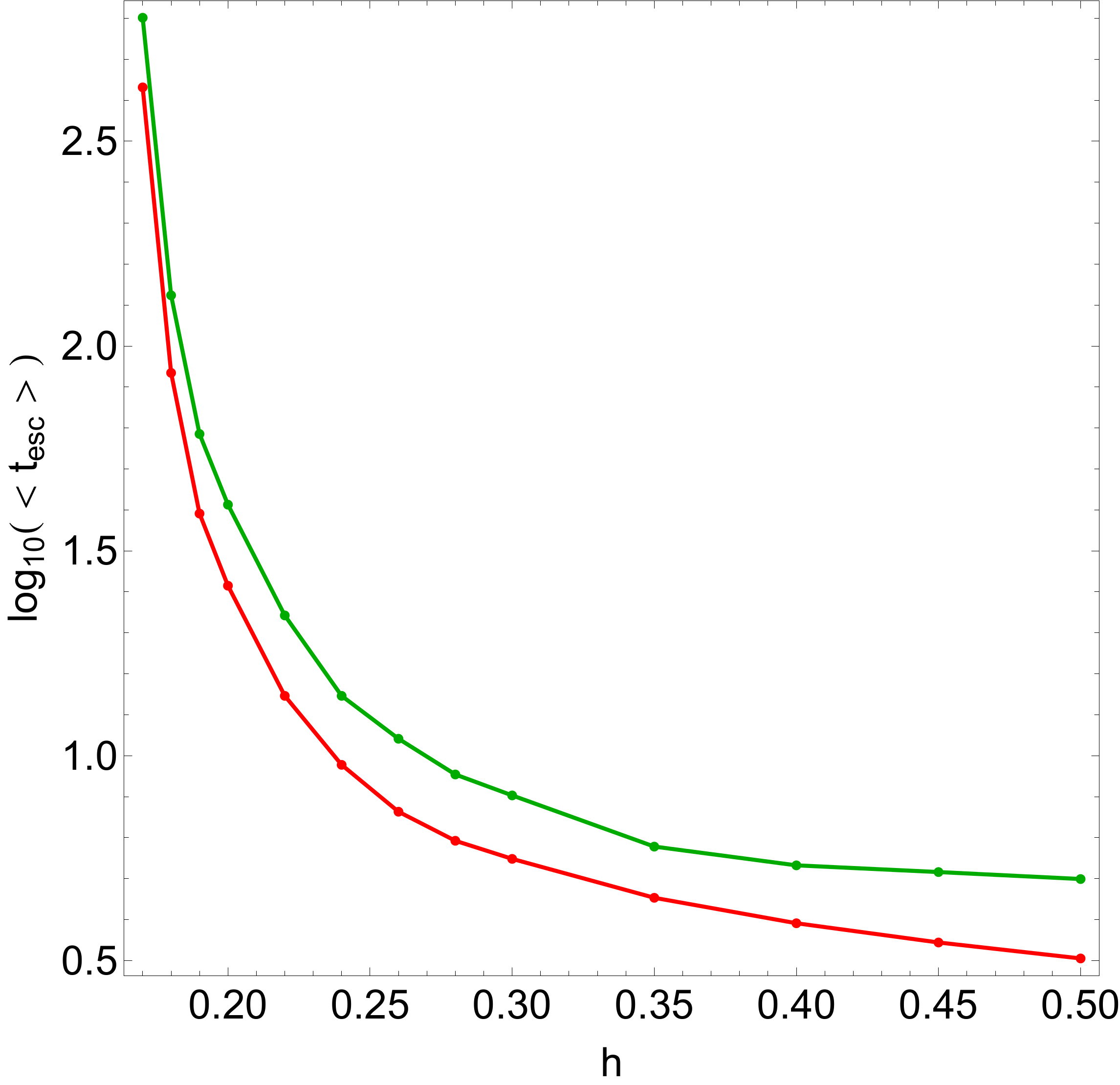}
\end{center}
\caption{Evolution of the logarithm of the average escape time of the orbits, $\log_{10}\left( < t_{\rm esc} > \right)$, as a function of the total orbital energy $h$, for both the configuration $(x,y)$ space (green) and the phase $(y\dot{y})$ space (red).}
\label{tavg}
\end{figure}

We continue our exploration of the escape process in the phase $(y,\dot{y})$ space. The structure of the $(y,\dot{y})$ phase plane for the same set of values of the energy is shown in Fig. \ref{ypy}. We observe a similar behavior to that discussed for the configuration $(x,y)$ plane in Fig. \ref{xy}. The outermost black solid line is the limiting curve which is defined as
\begin{equation}
f(y,\dot{y}) = \frac{1}{2}\dot{y}^2 + V(x = 0, y) = h.
\label{zvc}
\end{equation}
Here we must point out, that this $(y,\dot{y})$ phase plane is not a classical Poincar\'{e} Surface of Section (PSS), simply because escaping orbits in general, do not intersect the $x = 0$ axis after a certain time, thus preventing us from defining a recurrence time\footnote{In a classical PSS a recurrence time is defined as the time interval between two successive orbit intersections with a particular axis (i.e. the $x = 0$).}. A classical Poincar\'{e} surface of section exists only if orbits intersect an axis, like $x = 0$, at least once within a certain time interval. Nevertheless, in the case of escaping orbits we can still define local surfaces of section which help us to understand the orbital behavior of the dynamical system.

Once more, we can distinguish in the phase plane fractal regions where we cannot predict the particular escape channel and regions occupied by escape basins. These basins are either broad well-defined regions, or elongated bands of complicated structure spiralling around the center. We see that again for values of energy close to the escape energy there is a considerable amount of trapped orbits and the degree of fractalization of the phase plane is high. As we proceed to higher energy levels however, the rate of trapped orbits reduces, the phase plane becomes less and less fractal and is occupied by well-defined basins of escape. The distribution of the escape times $t_{\rm esc}$ of orbits on the $(y,\dot{y})$ plane is shown in Fig. \ref{ypyt}. It is evident, that orbits with initial conditions inside the exit basins escape from the system very quickly, or in other words, they possess extremely small escape periods. On the contrary, orbits with initial conditions located in the fractal parts of the phase plane need considerable amount of time in order to escape. It is seen in Fig. \ref{ypy} that for $h > 0.25$ all the Kolmogorov-Arnold-Moser (KAM) regime vanishes \cite{BBS08} and therefore, all the initial conditions of orbits escape through one of the exits.

It would be very interesting to monitor the evolution of the average value of the escape time $< t_{\rm esc} >$ of the orbits as a function of the total orbital energy $h$, for both the configuration $(x,y)$ as well as the phase $(y,\dot{y})$ space. Our results are presented in Fig. \ref{tavg}(a-b), where the values of the escape time are given in logarithmic scale. For both cases the average escape time of the orbits were obtained from the data of Figs. \ref{xy} and \ref{ypy}. We observe that for low energy levels the average escape period of orbits is more than 400 time units. However as the value of the energy increases the escape time of the orbits reduces rapidly. If we want to justify the behaviour of the escape time we should take into account the geometry of the open ZVCs. In particular, as the total orbital energy increases the three symmetrical escape channels become more and more wide and therefore, the test particles need less and less time until they find one of the four symmetrical openings in the ZVC and escape to infinity. This geometrical feature explains why for low values of the energy orbits consume large time periods wandering inside the open ZVC until they eventually locate one of the exits and escape to infinity.

\subsubsection{An overview analysis}
\label{over}

\begin{figure*}[!t]
\centering
\resizebox{\hsize}{!}{\includegraphics{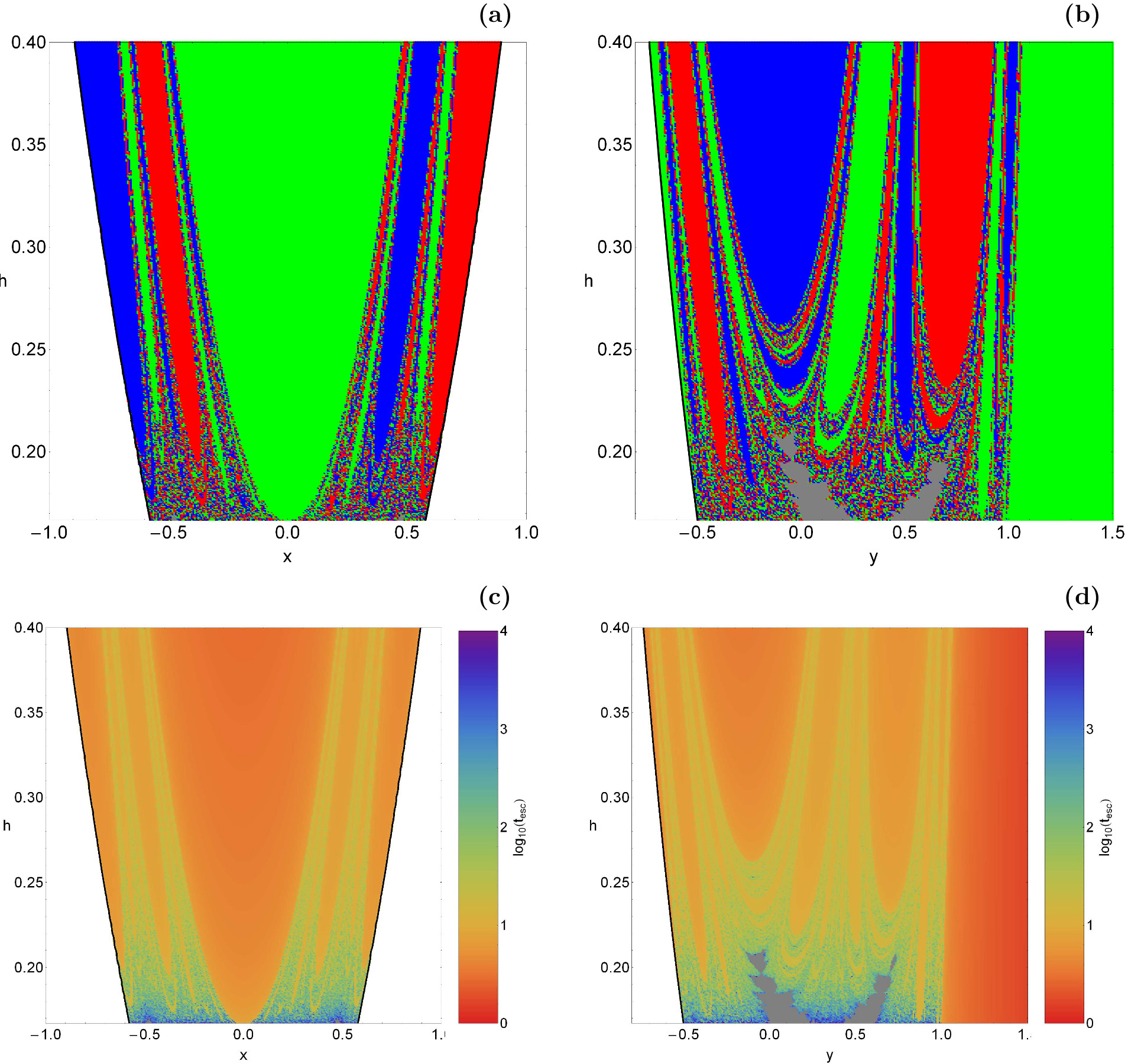}}
\caption{Orbital structure of the (a-upper left): $(x,h)$ plane; (b-upper right): $(y,h)$ plane when $h \in (1/6, 0.4]$. These diagrams provide a detailed analysis of the evolution of the trapped and escaping orbits when the parameter $h$ changes. The color code is the same as in Fig. \ref{xy}; (c-d): the distribution of the corresponding escape times of the orbits. In this type of grid representation the stability islands of regular orbits which are indicated by gray color can be identified more easily.}
\label{xyht}
\end{figure*}

The color-coded grids in configuration $(x,y)$ as well as in the phase $(y,\dot{y})$ plane space provide sufficient information on the phase space mixing however, for only a fixed value of the energy integral. H\'{e}non back in the late 60s \cite{H69}, introduced a new type of plane which can provide information not only about stability and chaotic regions but also about areas of bounded and unbounded motion using the section $y = \dot{x} = 0$, $\dot{y} > 0$ (see e.g., \cite{BBS08}). In other words, all the orbits of the test particles are launched from the $x$-axis with $x = x_0$, parallel to the $y$-axis. Consequently, in contrast to the previously discussed types of planes, only orbits with pericenters on the $x$-axis are included and therefore, the value of the energy $h$ can be used as an ordinate. In this way, we can monitor how the energy influences the overall orbital structure of our dynamical system using a continuous spectrum of energy values rather than few discrete energy levels. In Fig. \ref{xyht}a we present the orbital structure of the $(x,h)$ plane when $h \in (1/6,0.4]$, while in Fig. \ref{xyht}c the distribution of the corresponding escape time of orbits is depicted. The boundaries between bounded and unbounded motion are now seen to be more jagged than shown in the previous types of grids. In addition, we found in the blow-ups of the diagram many tiny islands of stability\footnote{From chaos theory we expect an infinite number of islands of (stable) quasi-periodic (or small scale chaotic) motion (see e.g., \cite{BGO90,O93}).}. It is evident that for low values of the energy $(h < 0.2)$ the outer parts of the $(x,h)$ plane exhibit a high degree of fractality which however is considerably reduced at higher energy levels where well-formed basins of escape dominate the plane.

In order to obtain a more complete view of the orbital structure of the system, we follow a similar numerical approach to that explained before but in this case we use the section $x = \dot{y} = 0$, $\dot{x} > 0$, considering orbits that are launched from the $y$-axis with $y = y_0$, parallel to the $x$-axis. This allow us to construct again a two-dimensional (2D) plane in which the $y$ coordinate of orbits is the abscissa, while the value of the energy $h$ is the ordinate. Fig \ref{xyht}b shows the structure of the $(y,h)$ plane, while the distribution of the corresponding escape time of orbits is given in Fig. \ref{xyht}d. We see, that for low values of the energy close to the escape energy, there is a considerable amount of trapped orbits inside stability regions surrounding by a highly fractal structure. This pattern however changes for larger energy levels $(h > 0.22)$, where there are no trapped regular orbits and the vast majority of the grid is covered by well-formed basins of escape, while fractal structure is confined only near the boundaries of the escape basins.

\begin{figure*}[!t]
\centering
\resizebox{\hsize}{!}{\includegraphics{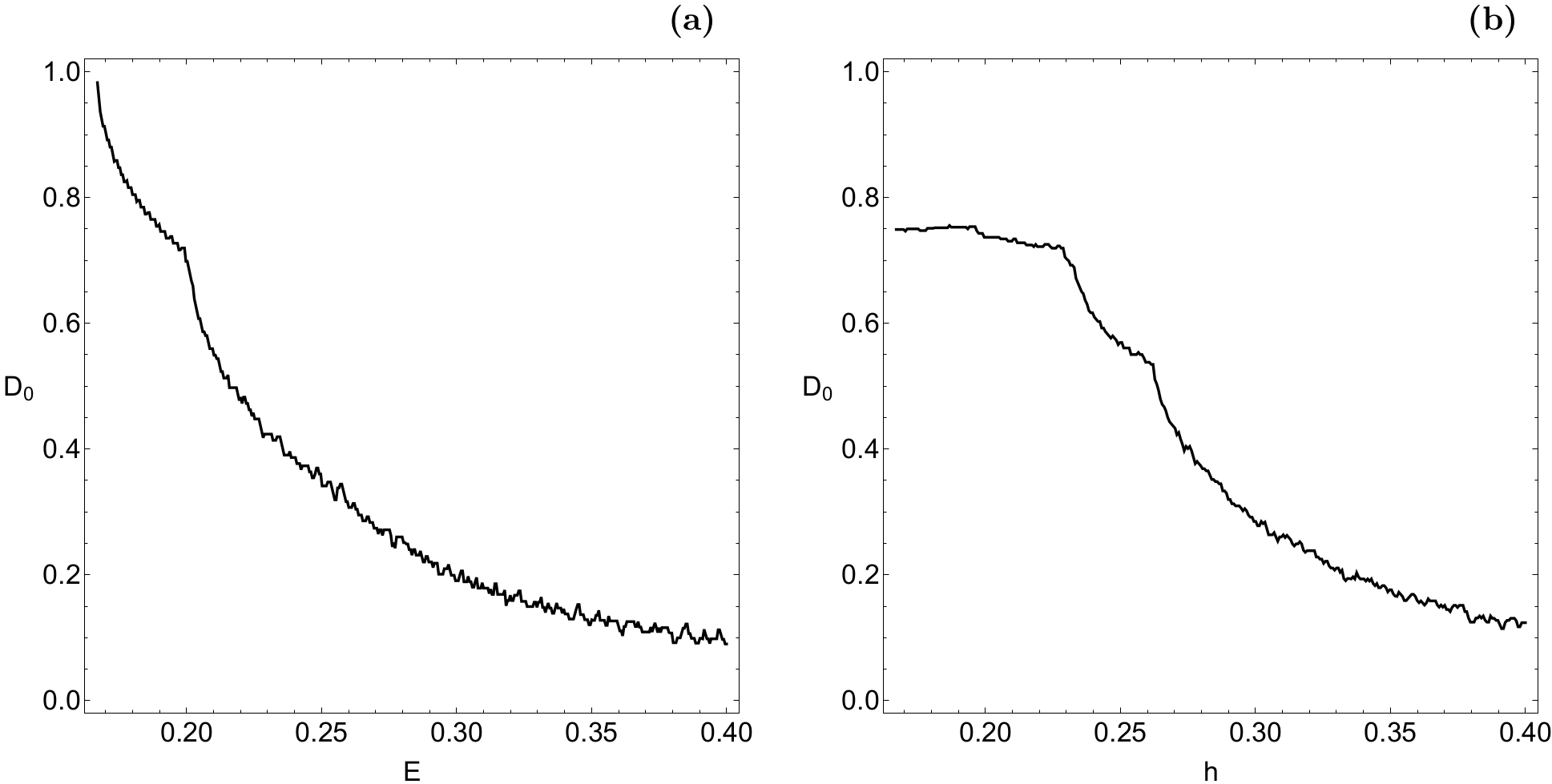}}
\caption{Evolution of the fractal dimension $D_0$ of the (a-left): $(x,h)$-plane and (b-right): $(y,h)$-plane of Figs. \ref{xyht} as a function of the total energy $h$. $D_0 = 1$ means total fractality, while $D = 0$ implies zero fractality.}
\label{frac}
\end{figure*}

In the previous two subsections we discuss fractality of the phase space in a qualitative way. In particular, rich and highly fractal domains are those in which we cannot predict through which exit channel the particle will escape since the particle chooses randomly an exit. On the other hand, inside the escape basins where the degree of fractality is zero the escape process of the particles is well known and predictable. At this point, we shall provide a quantitative analysis of the degree of fractality for the grids shown in Figs. \ref{xyht}. In order to measure the fractality we have computed the uncertainty dimension \cite{O93} for different values of the total energy following the computational method introduced in \cite{AVS01}. Obviously, this quantity is independent of the initial conditions used to compute it. The way to do it is the following. We calculate the exit for certain initial condition $(x,h)$ and $(y,h)$. Then, we compute the exit for the initial conditions $(x - \epsilon, h)$, $(x + \epsilon, h)$ and $(y - \epsilon, h)$, $(y + \epsilon, h)$ for a small $\epsilon$ and if all of them coincide, then this point is labeled as ``certain". If on the other hand they do not, it will be labeled as ``uncertain". We repeat this procedure for different values of $\epsilon$. Then we calculate the fraction of initial conditions that lead to uncertain final states $f(\epsilon)$. There exists a power law between $f(\epsilon)$ and $\epsilon$, $f(\epsilon) \propto \epsilon^{\alpha}$, where $\alpha$ is the uncertainty exponent. The uncertainty dimension $D_0$ of the fractal set embedded in the initial conditions is obtained from the relation $D_0 = D - \alpha$, where $D$ is the dimension of the phase space. It is typical to use a fine grid of values of $x$ or $y$ and $h$ to calculate the uncertainty dimension.

The evolution of the uncertainty dimension $D_0$ when the energy is increased is shown in Fig. \ref{frac}(a-b). As it has just been explained, the computation of the uncertainty dimension is done for only a ``1D slice'' of initial conditions of Fig. \ref{xyht}, and for that reason $D_0 \in (0,1)$. It is remarkable that the uncertainty dimension tends to one, at least in the $(x,h)$ plane\footnote{For the $(y,h)$ plane we see that when the value of the energy tends to the energy of escape the uncertainty dimension tends to about 0.75, which is much lower with respect to what observe for the $(x,h)$ plane. This phenomenon is explained because at the outermost right side of the $(y,h)$ plane a basin of escape is always present thus reducing the degree of fractality in every energy level.}, when the energy tends to its minimum value $E_{esc} = 1/6$. This means that for that critical value of the energy of escape, there is a total fractalization of the $(x,h)$ space, and the chaotic set becomes ``dense" in the limit. Consequently, in this limit there are no smooth sets of initial conditions (see Fig. \ref{xyht}) and the only defined structures that can be recognized are the KAM-tori of quasi-periodic orbits. When the energy is increased however, the different smooth sets appear and tend to grow, while the fractal structures that coincide with the boundary between basins decrease. Finally for values of energy much greater than the energy of escape the uncertainty dimension tends to very low values ($D_0 \simeq 0.1$), but it never completely vanishes (zero fractality).

At this point, we would like to emphasize that there are many methods for computing the predictability of a dynamical system. Very recently a new tool, called ``basin entropy", has been developed for analyzing the uncertainty in dynamical systems \cite{DWG16}. This new qualitative method describes the notion of fractality and unpredictability in the context of basins of attraction or basins of escape. The basin entropy provides an excellent qualitative information regarding the fractality of the escape basins, so it would be very illuminating if we could determine how the basin entropy evolves as a function of the total orbital energy. In a future work we shall use this new tool in order to investigate, in several types of Hamiltonian systems, how the unpredictability is significantly being reduced as the energy increases. Furthermore, it would be very informative to compare the corresponding results derived from both the basin entropy and the uncertainty dimension.

The rich fractal structure of the $(x,h)$ and $(y,h)$ planes observed in Figs. \ref{xyht}(a-b) implies that the system has also a strong topological property, which is known as the Wada property \cite{AVS01}. The Wada property is a general feature of two-dimensional (2D) Hamiltonians with three or more escape channels. A basin of escape verifies the property of Wada if any initial condition that is on the boundary of one basin is also simultaneously on the boundary of three or even more escape basins (see e.g., \cite{BSBS12,KY91}). In other words, every open neighborhood of a point $x$ belonging to a Wada basin boundary has a nonempty intersection with at least three different basins. Hence, if the initial conditions of a particle are in the vicinity of the Wada basin boundary, we will not be able to be sure by which one of the three exits the orbit will escape to infinity. Therefore, if a Hamiltonian system verifies the property of Wada, the unpredictability is even stronger than if it only had fractal basin boundaries \cite{BGOB88,O93}. If an orbit starts close to any point in the boundary, it will not be possible to predict its future behavior, as its initial conditions could belong to any of the other escape basins. This special topological property has been identified and studied in several dynamical systems (see e.g., \cite{AVS09,KY91,PCOG96}) and it is a typical property in open Hamiltonian systems with three or more escape channels.

\begin{figure*}[!t]
\resizebox{\hsize}{!}{\includegraphics{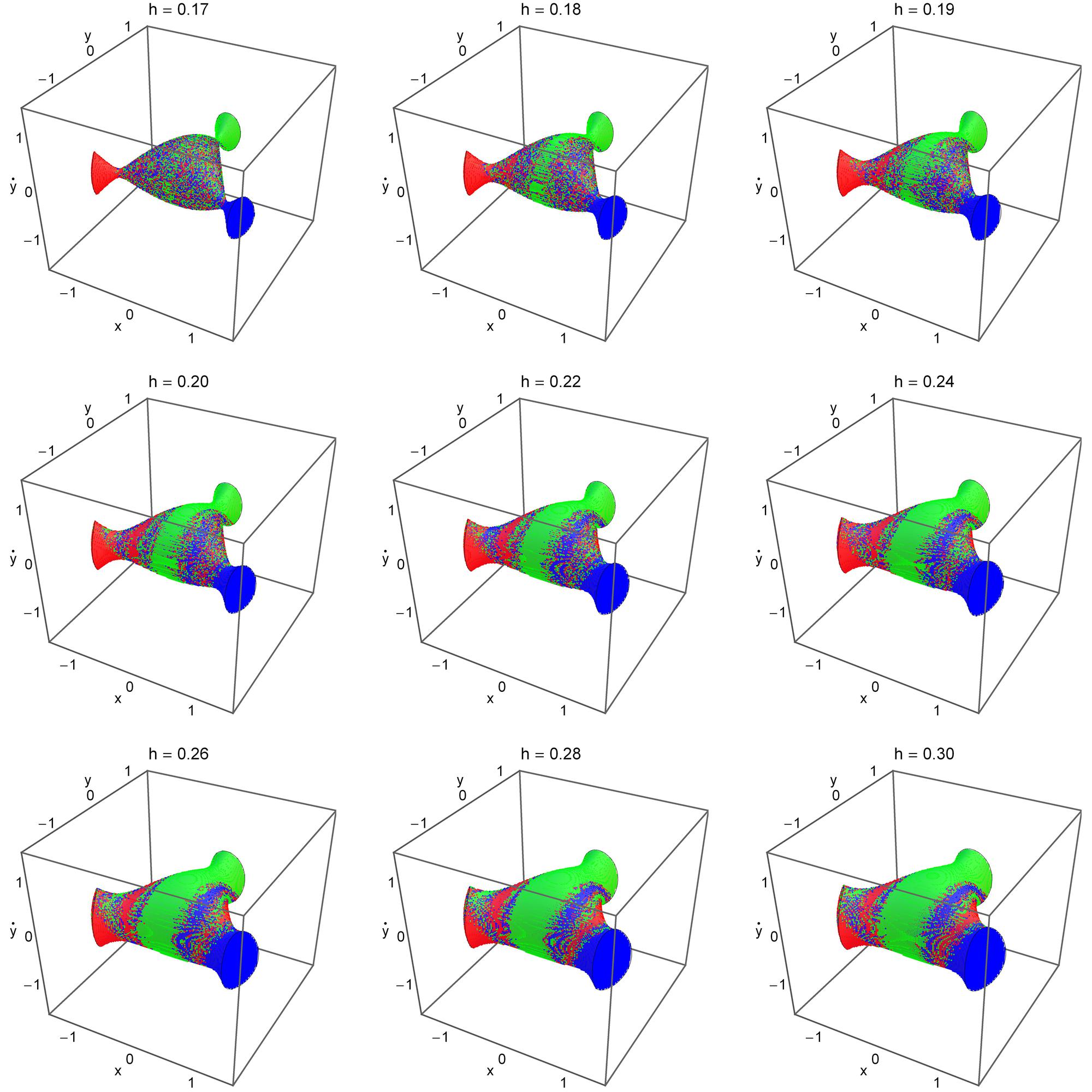}}
\caption{Orbital structure of three dimensional distributions of initial conditions of orbits in the 3D $(x,y,\dot{y})$ subspace for several values of the energy $E$. The color code is the same as in Fig. \ref{xy}.}
\label{3d}
\end{figure*}

\begin{figure*}[!t]
\resizebox{\hsize}{!}{\includegraphics{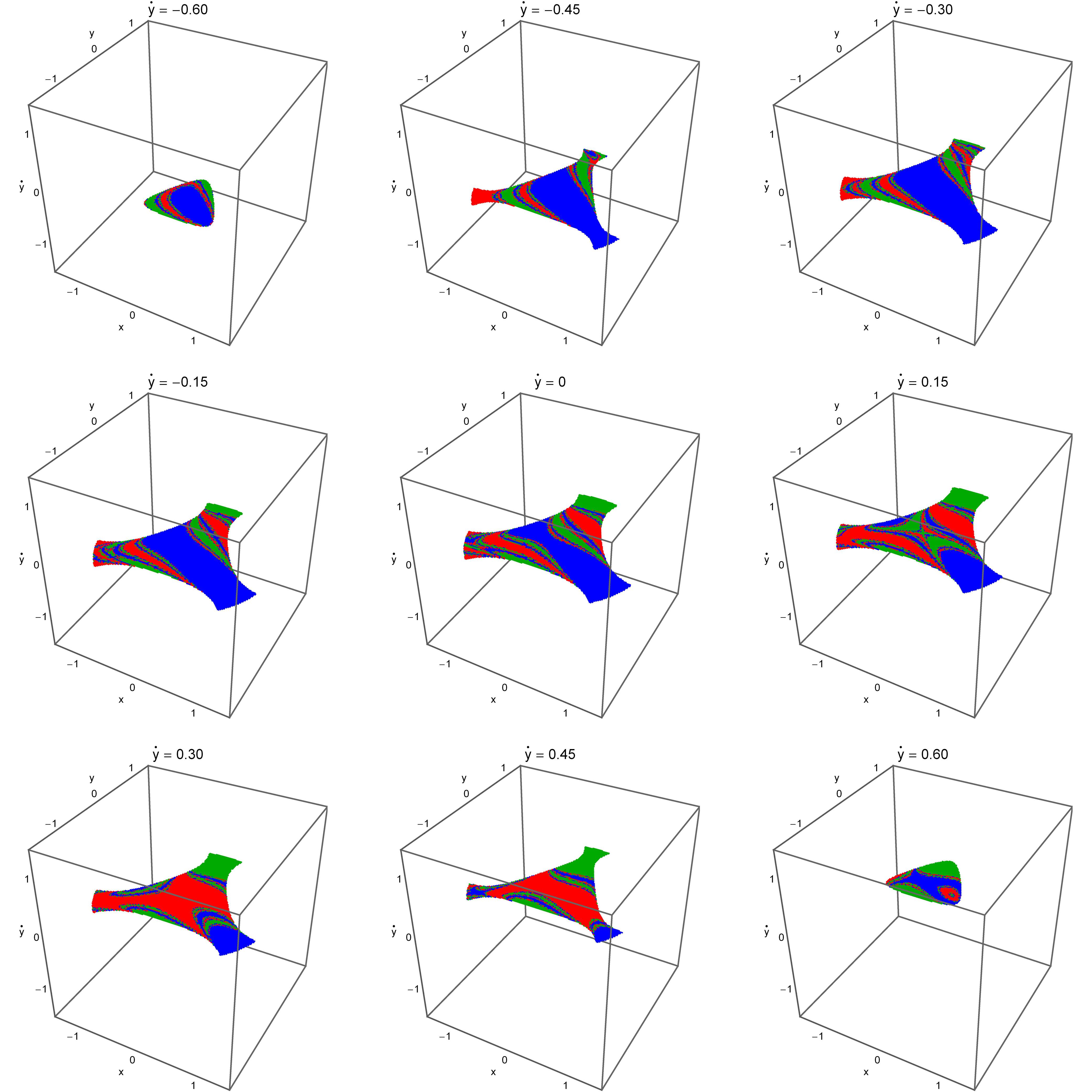}}
\caption{A tomographic view of the solid grid of Fig. \ref{3d} when $h = 0.30$ showing slices on the configuration $(x, y)$ plane when $\dot{y} = (-0.60, -0.45, -0.30, -0.15, 0, 0.15, 0.30, 0.45, 0.60)$.}
\label{tomo}
\end{figure*}

\subsection{Results for the 3D space}

In the previous subsections we investigated the escape dynamics of orbits using two dimensional grids of initial conditions in several types of planes (or in other words, in several 2D subspaces of the whole 4D\footnote{The dimension of the complete phase space of a Hamiltonian system with $N$ degrees of freedom is $2N$. This phase space is foliated into $2N - 1$ dimensional invariant leaves corresponding to the numerical values of the Hamiltonian. However, the dimension of the entire phase space always remains $2N$, irrelevantly of any possible invariant foliations.} phase space). In this subsection we will expand our numerical exploration using three dimensional distributions of initial conditions of orbits. Being more precise, for a particular value of the energy we define inside the corresponding zero velocity surface uniform grids of initial conditions $(x_0,y_0,\dot{y_0})$, while the initial value of $\dot{x} > 0$ is always obtained from the energy integral of motion (\ref{ham}).

In Fig. \ref{3d} we present the orbital structure of the three dimensional distributions of initial conditions of orbits in the $(x,y,\dot{y})$ subspace for the same set of values of the energy. The color code is the same as in Fig. \ref{xy}. At this point, we should emphasize that in this case the distribution of the initial conditions of the orbits cannot illustrate any longer the $2\pi/3$ symmetry of the system, which is only visible in polar coordinates and only in the two-dimensional $(x,y)$ plane. This means that the three escape channels of the potential are not equiprobable.

It is seen in Fig. \ref{3d} that the grids of the initial conditions of the orbits are in fact three dimensional solids and therefore only their outer surface is visible. However, a tomographic-style approach can be used in order to penetrate and examine the interior region of the solids (e.g., \cite{Z14a}). According to this method, we can plot two dimensional slices of the solid grid by defying specific levels to a primary plane (i.e., the $(x,y)$ plane). Fig. \ref{tomo} shows the evolution of the structure of the configuration $(x, y)$ plane when $\dot{x}$ = (-0.60, -0.45, -0.30, -0.15, 0, 0.15, 0.30, 0.45, 0.60) for $h = 0.30$. We see that the structure evolves rapidly and non-uniformly (constant interplay between escape basins and fractal structure) thus implying that the escape process in this dynamical system is, by all means, a very complex and fascinating procedure. More precisely, we observe that for low values of the velocity $\dot{y}$ escaping orbits through channel 3 dominate, while for high values of $\dot{y}$ (around 0.30 to 0.45) channel 2 seems to be more preferable. Similar tomographic plots can be obtained also for the other values of the energy presented in Fig. \ref{3d}.

Taking into account that the three escape channels are no longer equiprobable (this also applies for the $(y,\dot{y})$ space) we could present the evolution of the percentages of all types of orbits as a function of the value of the total orbital energy $h$. However we feel that this could be very confusing or even misleading regarding the physics behind the problem. Once more we have to emphasize that the $2\pi/3$ symmetry is not broken in the 3D subspace. However, in this case, due to the particular choice of the initial conditions of the orbits, and not because of some internal property of the system, the symmetry cannot be illustrated.

Before closing this section we would like to mention that our numerical computations suggest that the average escape time of the orbits, with initial condition in the three-dimensional $(x,y,\dot{y})$ space, exhibits a similar evolution to that discussed earlier in Fig. \ref{tavg} for the distribution of orbits in both the configuration and the phase space.

\section{Discussion}
\label{disc}

The aim of this work was to present a complete view of the escape dynamics in the classical H\'{e}non-Heiles Hamiltonian. This dynamical system has the key feature of having a finite energy of escape. In particular, for energies smaller than the escape value, the equipotential surfaces are closed and therefore escape is impossible. For energy levels larger than the escape energy however, the equipotential surfaces open and several channels of escape appear through which the test particles are free to escape to infinity. Here we should emphasize, that if a test particle has energy larger than the escape value, this does not necessarily mean that the test particle will certainly escape from the system and even if escape does occur, the time required for an orbit to cross an unstable Lyapunov orbit and hence escape to infinity may be very long compared with the natural crossing time.

For the numerical integration of the orbits in each type of plane, we needed roughly between 1 minute and 2 days of CPU time on a Pentium Dual-Core 2.2 GHz PC, depending both on the amount of trapped orbits and on the escape rates of orbits in each case. For each initial condition, the maximum time of the numerical integration was set to be equal to $10^5$ time units however, when a test particle escapes the numerical integration is effectively ended and proceeds to the next initial condition.

This present paper is a combination of already known results and also new outcomes (see e.g, Figs. \ref{xyt}, \ref{ypyt}, \ref{frac}, \ref{3d}, \ref{tomo}) regarding the H\'{e}non-Heiles system. In particular, as far as we know, this is the first time that a systematic classification of sets of initial conditions inside the 3D $(x,y,\dot{y})$ subspace takes place and therefore the results presented in Figs. \ref{3d} and \ref{tomo} are, without any doubt, the most novel ones of this investigation. We hope that the present analysis and the corresponding numerical outcomes to provide to the readers a complete overview regarding the complicated escape dynamics of the H\'{e}non-Heiles Hamiltonian system.

It is in our future plans to investigate in more detail the parametric evolution of the degree of fractality as a function of the total orbital energy by using, and therefore comparing, the numerical results derived from the computation of both the uncertainty dimension and the basin entropy. Furthermore, it would be of particular interest to determine how the degree of fractality is influenced (or not) by the total number of the exits of the Hamiltonian systems.

\section*{Acknowledgments}

The author would like to express his warmest thanks to the two anonymous referees for the careful reading of the manuscript and for all the apt suggestions and comments which allowed us to improve both the quality and the clarity of the paper.

\section*{Compliance with ethical standards}

{\bf{Conflict of interest}}: The author declares that he has no conflict of interest.

\end{document}